\RequirePackage[hyphens]{url}
\documentclass{article}
\usepackage{authblk}
\usepackage[margin=1.4in]{geometry}

\usepackage{xspace}
\usepackage{graphicx}
\usepackage{multirow}
\usepackage{hhline}
\usepackage{comment}
\usepackage{adjustbox}
\usepackage{xcolor}
\usepackage{pifont}
\usepackage{subfigure}
\usepackage{tabulary}
\usepackage{array}
\usepackage{colortbl}
\usepackage{color,soul}
\usepackage[colorlinks=true,
citecolor=blue,
filecolor=black,
linkcolor=blue,
urlcolor=blue]{hyperref}
\usepackage{booktabs} 

\renewcommand{\baselinestretch}{1.0}
\usepackage[switch,columnwise]{lineno} 

\usepackage{amsmath}
\usepackage{amsfonts}       
\usepackage{dsfont}

\newcommand\extrafootertext[1]{%
    \bgroup
    \renewcommand\thefootnote{\fnsymbol{footnote}}%
    \renewcommand\thempfootnote{\fnsymbol{mpfootnote}}%
    \footnotetext[0]{#1}%
    \egroup
}
\usepackage{listings}
\usepackage{pythonhighlight}

\usepackage{xcolor}

\definecolor{codegreen}{rgb}{0,0.6,0}
\definecolor{codegray}{rgb}{0.5,0.5,0.5}
\definecolor{codepurple}{rgb}{0.58,0,0.82}
\definecolor{backcolour}{rgb}{0.95,0.95,0.92}

\lstdefinestyle{mystyle}{
    backgroundcolor=\color{backcolour},
    commentstyle=\color{codegreen},
    keywordstyle=\color{magenta},
    numberstyle=\tiny\color{codegray},
    stringstyle=\color{codepurple},
    basicstyle=\ttfamily\footnotesize,
}

\title{DeepSpeed-FastGen: High-throughput Text Generation for LLMs via MII and DeepSpeed-Inference}
\author[]{Connor Holmes, Masahiro Tanaka, Michael Wyatt, Ammar Ahmad Awan, Jeff Rasley, Samyam Rajbhandari, Reza Yazdani Aminabadi, Heyang Qin, Arash Bakhtiari, Lev Kurilenko, Yuxiong He}
\affil[]{Microsoft DeepSpeed (www.deepspeed.ai)}

\date{}

\begin{document}

\maketitle

\begin{abstract}

The deployment and scaling of large language models (LLMs) have become critical as they permeate various applications, demanding high-throughput and low-latency serving systems. Existing frameworks struggle to balance these requirements, especially for workloads with long prompts. This paper introduces DeepSpeed-FastGen, a system that employs Dynamic SplitFuse, a novel prompt and generation composition strategy, to deliver up to 2.3x higher effective throughput, 2x lower latency on average, and up to 3.7x lower (token-level) tail latency, compared to state-of-the-art systems like vLLM. We leverage a synergistic combination of DeepSpeed-MII and DeepSpeed-Inference to provide an efficient and easy-to-use serving system for LLMs. DeepSpeed-FastGen's advanced implementation supports a range of models and offers both non-persistent and persistent deployment options, catering to diverse user scenarios from interactive sessions to long-running applications. We present a detailed benchmarking methodology, analyze the performance through latency-throughput curves, and investigate scalability via load balancing. Our evaluations demonstrate substantial improvements in throughput and latency across various models and hardware configurations. We discuss our roadmap for future enhancements, including broader model support and new hardware backends. The DeepSpeed-FastGen code is readily available for community engagement and contribution.

\end{abstract}

\section{Introduction}
Large language models (LLMs) like GPT-4~\cite{gpt4} and LLaMA~\cite{llama} have emerged as a dominant workload in serving a wide range of applications infused with AI at every level.
From general chat models to document summarization, and from autonomous driving to copilots at every layer of the software stack, the demand to deploy and serve these models at scale has skyrocketed.
While frameworks like DeepSpeed, PyTorch~\cite{pytorch}, and several others can regularly achieve good hardware utilization during LLM training, the interactive nature of these applications and the poor arithmetic intensity of tasks like open-ended text generation have become the bottleneck for inference throughput in existing systems.

To this end, frameworks like vLLM~\cite{vllm} powered by PagedAttention and research systems like Orca~\cite{orca} have significantly improved the performance of inference for LLMs.
However, these systems still struggle to provide consistent quality of service, particularly for workloads with longer prompts.
These long prompt workloads are becoming increasingly important as more and more models, like MPT-StoryWriter~\cite{mptstorywriter}, and systems, such as DeepSpeed Ulysses~\cite{deepspeedulysses}, support context windows stretching to tens of thousands of tokens.
To better understand the problem space, we provide detailed examples of how text generation works for LLMs in two distinct phases called prompt processing and generation. When systems treat them as distinct phases, generation will be preempted by prompt processing that risks breaking the service level agreements (SLAs).

Today, we are glad to present DeepSpeed-FastGen, a system that overcomes these limitations by leveraging the proposed Dynamic SplitFuse technique and offers up to 2.3x higher effective throughput, 2x lower latency on average, and up to 3.7x lower (token-level) tail latency, compared to state-of-the-art systems like vLLM.
DeepSpeed-FastGen leverages the combination of DeepSpeed-MII and DeepSpeed-Inference to provide an easy-to-use serving system.

\label{sec:introduction}

\section{Existing LLM Serving Techniques in Literature}
\label{sec:easy-to-use}
A text generation workload for a single sequence consists of two phases: 1) prompt processing, in which the user-provided text is efficiently processed as a batch of tokens to build a key-value (KV) cache for attention, and 2) token generation, which will add a single token to that cache and generate a new token. Over the course of generating a sequence of text, the model will make many forward calls to the model to generate the full sequence of text. Two major techniques have been proposed in the literature and deployed in systems that address various limitations and bottlenecks that may arise during these phases.

\subsection{Blocked KV Caching}

vLLM identified that memory fragmentation due to large monolithic KV-caches significantly reduced the concurrency of LLM serving systems and proposed Paged Attention~\cite{pagedattention} to enable non-contiguous caches and increase total system throughput. Rather than assign individual variable-sized contiguous chunks of memory, the underlying storage in the KV cache is fixed-sized blocks (also known as pages). The blocked KV-cache increases system throughput by increasing the amount of potential sequence concurrency by eliminating KV-cache induced memory fragmentation. Non-contiguous KV cache implementations are also included in HuggingFace TGI~\cite{huggingfacetgi} and NVIDIA TensorRT-LLM~\cite{tensorrtllm}.

\subsection{Continuous Batching}
In the past, dynamic batching, in which a server would wait for multiple requests to process in phase with each other, was used to improve GPU utilization. However, this approach has drawbacks, as it typically requires padding inputs to identical lengths or stalling the system to wait to construct a larger batch.

Recent advancement in large language model (LLM) inference and serving has been focusing on fine granularity scheduling and optimizing memory efficiency. For instance, Orca proposes iteration-level scheduling (also known as continuous batching) which makes distinct scheduling decisions at each forward pass of the model. This allows requests to join/leave the batch as needed, eliminating the need for padding requests thus improving the overall throughput. In addition to Orca, continuous batching has been implemented in NVIDIA TRT-LLM, HuggingFace TGI, and vLLM.

In current systems, there are two primary approaches to implement continuous batching. In TGI and vLLM, the generation phase is preempted to perform prompt processing (called infill in TGI) before continuing with generation. In Orca, these phases are not distinguished; instead, Orca will add a prompt into the running batch so long as the total number of sequences doesn't reach a fixed bound. These approaches to varying degrees need to stall generation to process long prompts (see Section~\ref{splitfuse}).

We propose a novel prompt and generation composition strategy, Dynamic SplitFuse discussed at length in the next section.

\section{Dynamic SplitFuse: A Novel Prompt and Generation Composition Strategy}
\label{sec:full-rlhf}
DeepSpeed-FastGen is built to leverage continuous batching and non-contiguous KV caches to enable increased occupancy and higher responsivity for serving LLMs in the data center, similar to existing frameworks such as TRT-LLM, TGI, and vLLM. iIn order to achieve a new level of performance, DeepSpeed-FastGen introduces SplitFuse which leverages dynamic prompt and generation decomposition and unification to further improve continuous batching and system throughput.

\subsection{Three Performance Insights}

Before describing Dynamic SplitFuse, we answer three key performance questions that together motivate its design.

\subsubsection{What factors impact the forward pass of a single LLM?} 

In order to effectively schedule, it is necessary to understand what are the relevant independent variables the scheduling loop should control. We observe below that the composition of sequences in a forward pass (the batch size in sequences) has a negligible impact on performance compared to the raw number of tokens in the forward pass. This means an effective scheduler can be built around a single signal, the number of tokens in the forward pass.

\begin{figure}[htbp!]
\centering
\includegraphics[width=0.7\textwidth]{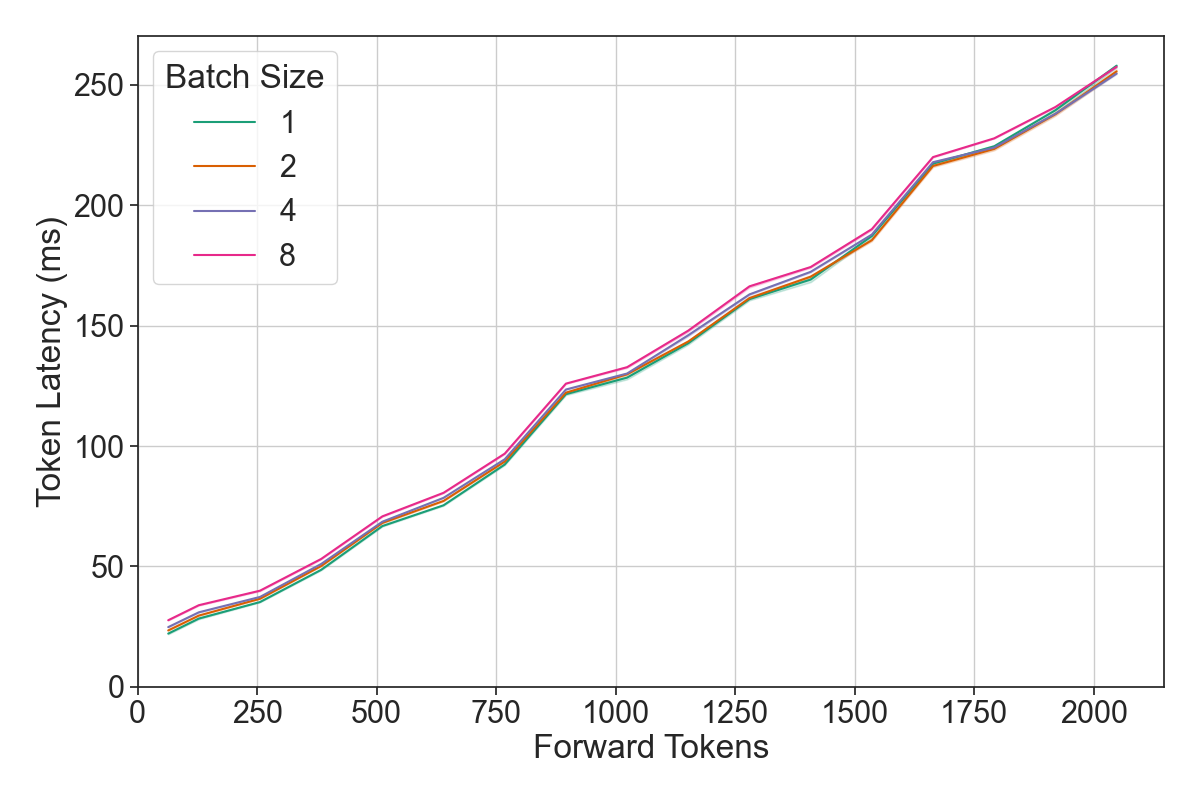}
\caption{Token latency (ms) is predominantly determined by the number of forward tokens rather than the batch sizes.}
\label{fig:observation_prompt_v_latency}
\end{figure}

\subsubsection{How does a model's throughput respond to changing the number of tokens in the forward pass?} 

An LLM has two key operating regions with a relatively steep transition. With a small number of tokens, the GPU bottleneck is reading the model from memory and so throughput scales with the number of tokens, whereas with many tokens the model is throughput bound by compute and sees near-constant throughput. The model should run highly efficiently if all forward passes are in the throughput-saturating region.

\begin{figure}[htbp!]
\centering
\includegraphics[width=0.7\textwidth]{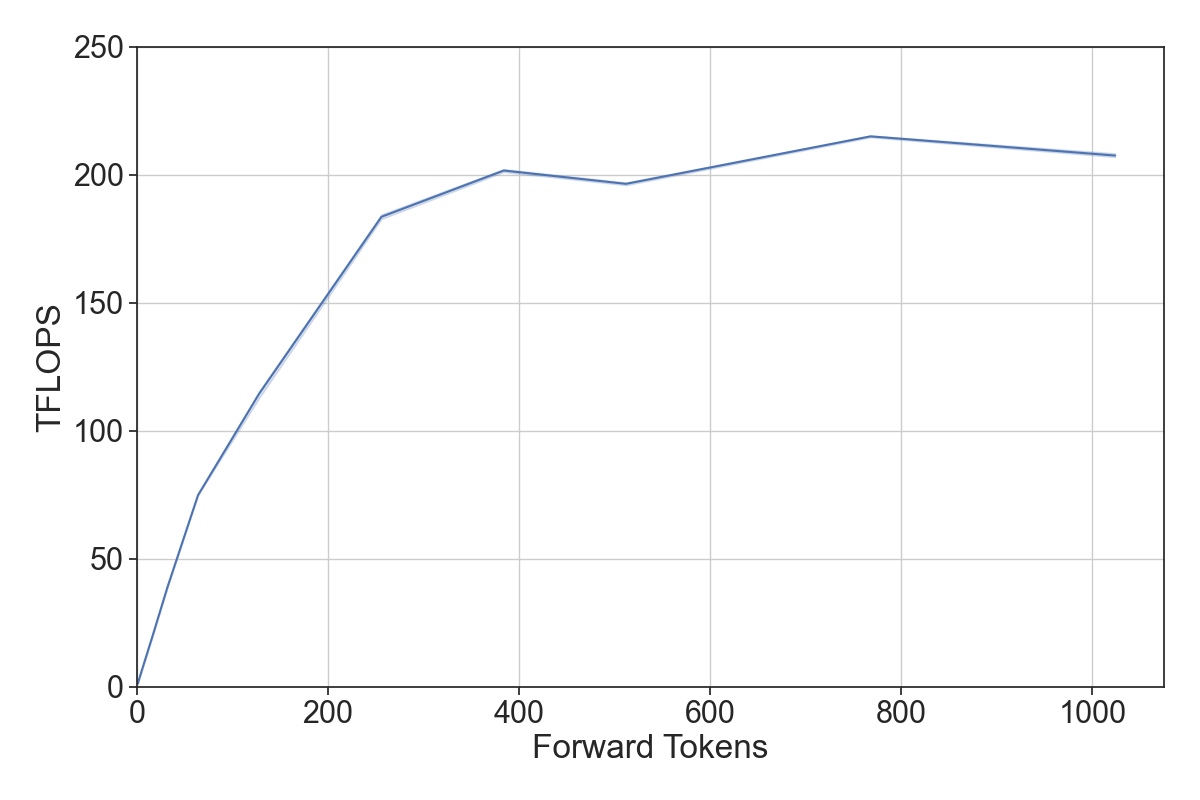}
\caption{The system reaches peak performance (throughput saturation region) as the number of tokens in the forward pass increase. Beyond that, near-constant throughput is observed.}
\label{fig:observation_prompt_v_flops}
\end{figure}

\subsubsection{How should a pool of tokens be scheduled across multiple forward passes?} We observe above that for well-aligned inputs the token-throughput curve is concave, which means the second derivative is bound to be less than or equal to 0. As an example, let $f(x)$ be a concave function of latency to throughput for a given model. For a concave function $f(x)$, the following holds:

  $$0 \geq \lim_{h \to 0} \frac{f(x + h) - 2f(x) + f(x - h)}{h^2}$$

  $$0 \geq f(x + h) - 2f(x) + f(x - h)$$

  $$2f(x) \geq f(x + h) + f(x - h)$$

This states that for a given pool of 2x tokens to process, the manner that maximizes throughput is that which evenly splits them between two batches. More generally, in a system that must consume and process P tokens over F forward passes, the ideal partitioning scheme will divide them equally.

\subsection{Dynamic SplitFuse} \label{splitfuse}

Dynamic SplitFuse is a novel token composition strategy for prompt processing and token generation. DeepSpeed-FastGen utilizes Dynamic SplitFuse to run at a consistent forward size by leveraging the capability to take partial tokens from prompts and compose this with generation. A similar approach has been proposed in Sarathi\cite{agrawal2023sarathi} where it splits a prompt into smaller chunks to combine more token generation with prompt processing and to run forward passes with consistent batch sizes.
In particular, Dynamic SplitFuse performs two key behaviors:

\begin{enumerate}
    \item Long prompts are decomposed into much smaller chunks and scheduled across multiple forward passes (iterations) with only the final pass performing any generation.
    \item Short prompts will be composed to exactly fill a target token budget. Even short prompts may be decomposed to ensure the budget is precisely met and the forward sizes are well-aligned.
\end{enumerate}

Together, these two techniques provide concrete benefits on all user metrics:

\begin{enumerate}
    \item Better Responsiveness: Since long prompts no longer require extremely long forward passes to process, the model will provide lower client latency. More forward passes are performed within the same window of time.
    \item Higher Efficiency: Fusion of short prompts to larger token budgets enables the model to consistently operate in the high throughput regime.
    \item Lower variance and better consistency: Since forward passes are of consistent size and forward pass size is the primary determinant of performance, the latency of each forward pass is much more consistent than competing systems as is the perceived generation frequency. There are no preemption or long-running prompts to increase the latency as in other prior work. This translates to a reduction of up to 3.7x P95 latency in generation as we show in Section \ref{sec:he}.
\end{enumerate}

\begin{figure}[htbp!]
\centering
\includegraphics[width=0.7\textwidth]{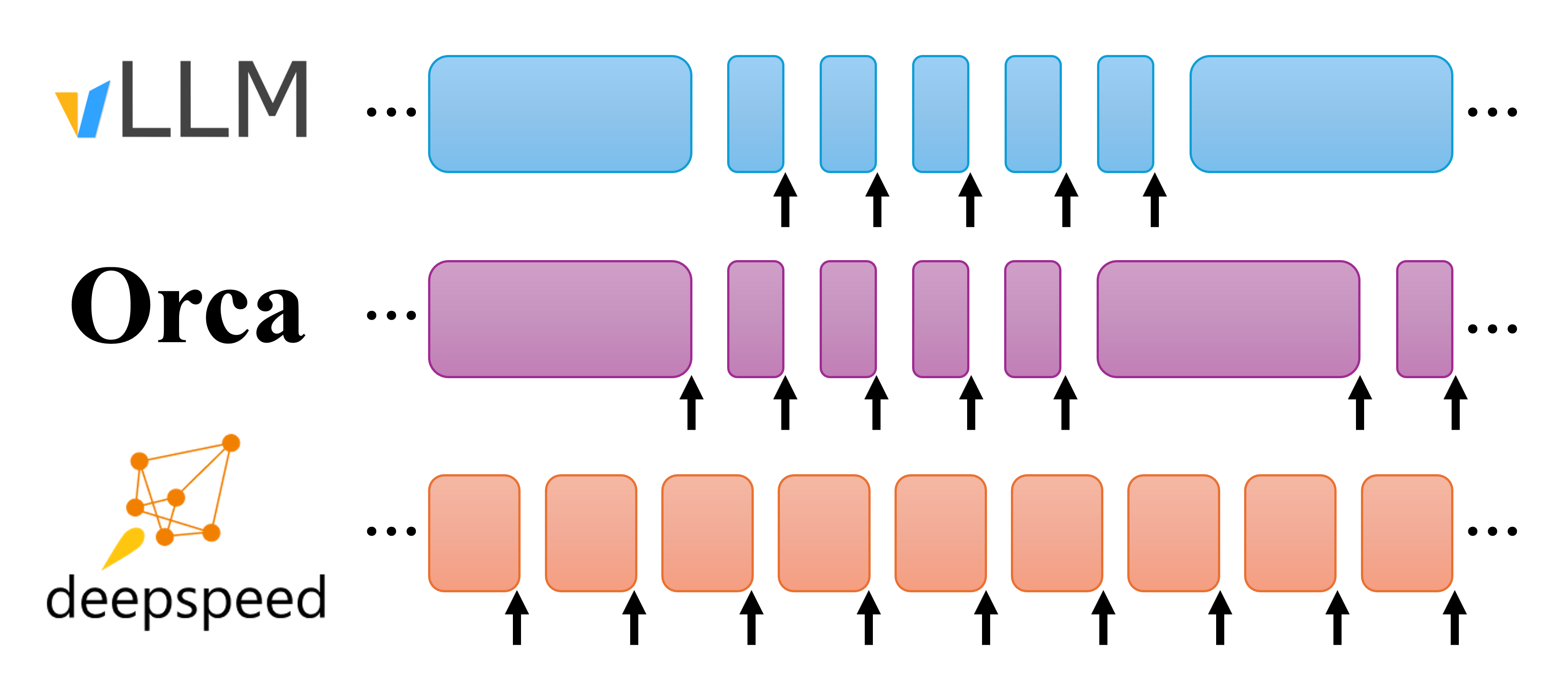}
\caption{Illustration of continuous batching strategies. Each block shows the execution of a forward pass.  An arrow indicates that the forward pass has sequences with one or more tokens generated. vLLM performs either token generations or prompt processing in a forward pass; token generation preempts prompt processing. Orca runs prompts at their complete length alongside generation. Dynamic SplitFuse performs dynamic composition of fixed-sized batches composed of both generation and prompt tokens.}
\label{fig:fastgen_overview_light}
\end{figure}

Consequently, DeepSpeed-FastGen will consume tokens from incoming prompts at a rate that permits fast ongoing generation while adding tokens to the system that increase system utilization, providing lower latency and higher throughput streaming generation to all clients as compared to other state-of-the-art serving systems.

\section{Performance Evaluation}
\label{sec:he}

DeepSpeed-FastGen provides state-of-the-art LLM serving performance leveraging its blocked KV cache and Dynamic SplitFuse continuous batching. We evaluate DeepSpeed-FastGen against vLLM~\cite{vllm} on a range of models and hardware configurations following the benchmarking methodology discussed below. The evaluation shows that DeepSpeed-FastGen achieves up to 2.3x higher effective throughput, 2x lower latency on average, and up to 3.7x lower (token-level) tail latency, compared to state-of-the-art systems like vLLM.

\subsection{Benchmarking Methodology}

We use two primary quantitative schemes for measuring performance.

\subsubsection{Throughput-Latency Curves} Two key metrics for production readiness are throughput (measured in requests per second) and latency (the responsiveness of each request). To measure this, we instantiate multiple clients (ranging from 1 to 32) concurrently and send requests (512 in total) to the server. The resulting latency of each request is measured at the endpoint and throughput is measured by the end-to-end time to complete the experiment.

\subsubsection{Effective Throughput} Interactive applications, such as chat applications, can have more stringent and complex requirements than can be captured by top-level metrics like end-to-end latency. In particular, we focus on the increasingly popular chat user scenario:

\begin{enumerate}
    \item A user initiates a task by sending a prompt.
    \item The system processes the prompt and returns the first token.
    \item Subsequent tokens are streamed to the user as they are produced.
\end{enumerate}

At each point in this process there is an opportunity for a system to provide an adverse user experience; for example, if the first token arrives too slowly or the generation appears to stop for some time. We propose an SLA framework that considers both of these dimensions.

As the lengths of prompts and generated texts vary significantly, affecting computational costs, it is impractical to set rigid SLA values for throughput and latency. Therefore, we define the SLA for prompt latency as \(\frac{\text{|tokens in prompt|}}{512}\) seconds (=\(512\) tokens/s). Additionally, considering humans' reading speed, we set the SLA for generation latency on the Exponential Moving Average (EMA) to 2, 4, or 6 tokens/sec. Requests that adhere to these SLAs are deemed successful, and the throughput of these successful requests is referred to as \textbf{effective throughput}.

We evaluate vLLM and DeepSpeed-FastGen on both Llama-2 7B, Llama-2 13B, and Llama-2 70B~\cite{llama} on NVIDIA A100, H100, and A6000.

\subsection{Throughput-Latency Analysis}

\begin{figure}
\centering
\includegraphics[width=\textwidth]{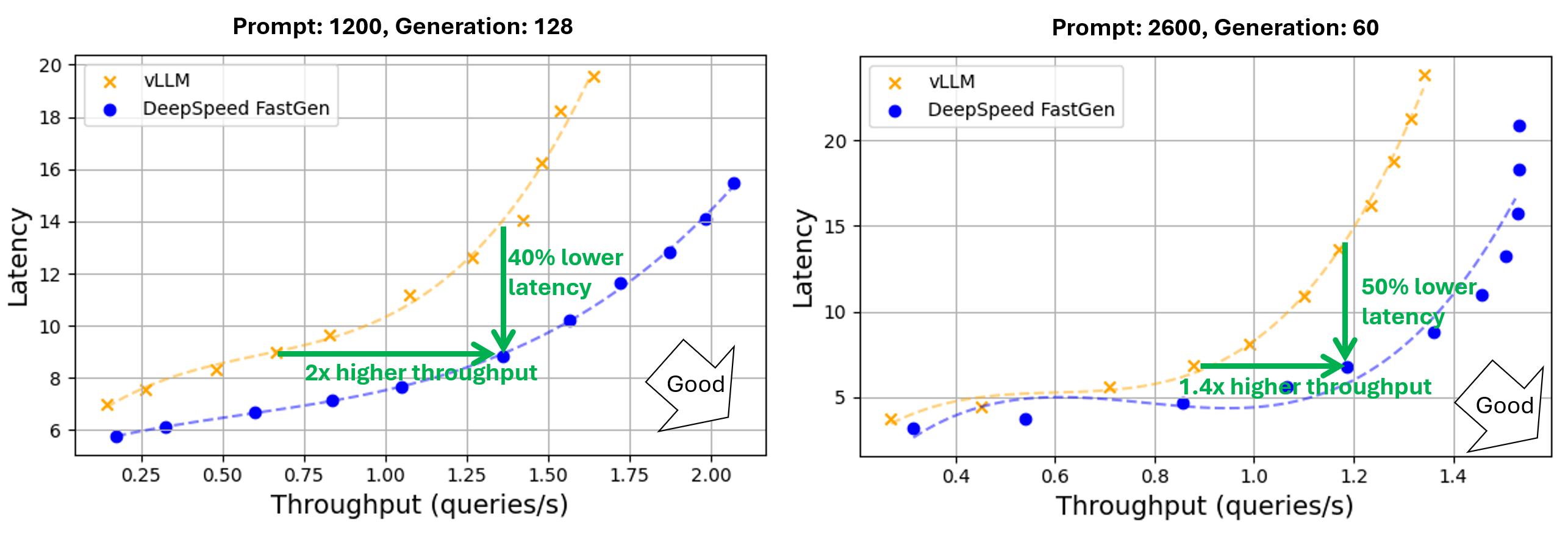}
\caption{Throughput and latency of text generation using Llama 2 70B (Tensor parallelism across 4 A100-80GB GPUs). A normal distribution was applied to prompt and generation lengths with averages of 1200/2600 and 128/60, respectively, and a 30\% variance.}
\label{fig:throughput_latency}
\end{figure}

\begin{figure}[ht]
\centering
\includegraphics[width=0.95\textwidth]{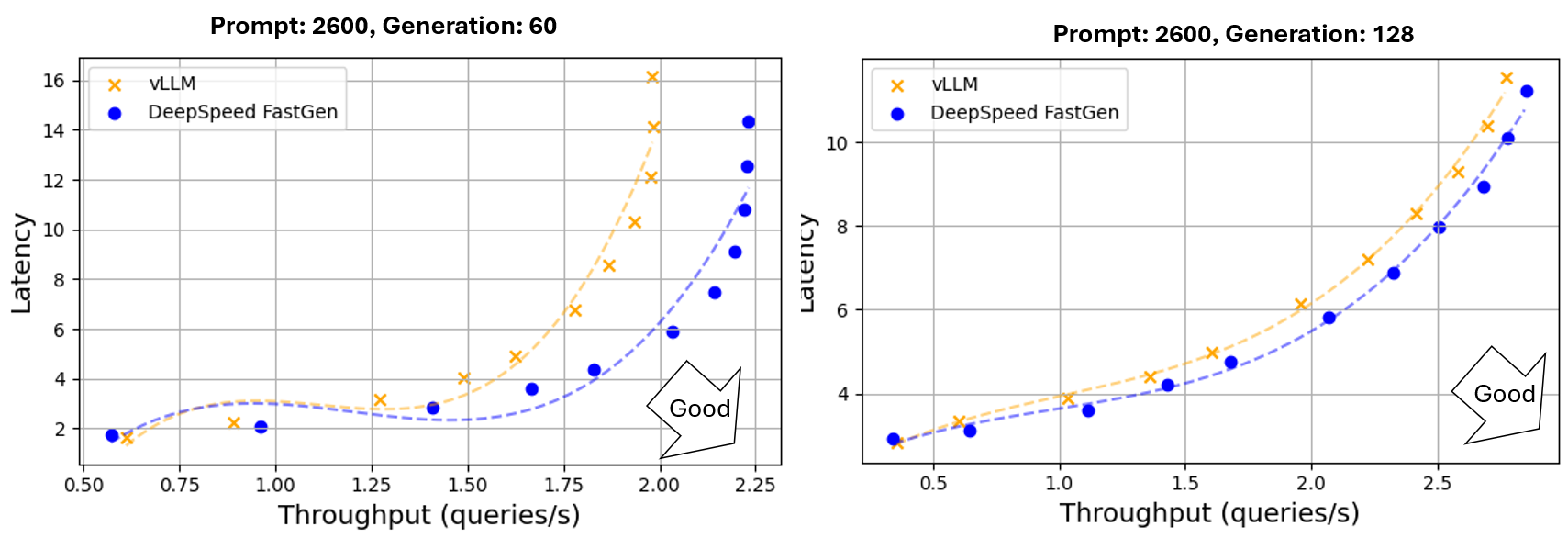}
\caption{Throughput and latency of text generation using Llama 2 13B (A100-80GB GPU, no tensor parallelism). A normal distribution was applied to prompt and generation lengths with averages of 1200/2600 and 60/128, respectively, and a 30\% variance.}
\label{fig:throughput_latency_13B}
\end{figure}

In this experiment, DeepSpeed-FastGen outperforms vLLM in both throughput and latency, providing equivalent latency with greater throughput or more responsive latency and the same throughput. On Llama-2 70B with 4 A100x80GB, DeepSpeed-FastGen demonstrates up to 2x higher throughput (1.36 rps vs. 0.67 rps) at identical latency (9 seconds) or up to 50\% latency reduction (7 seconds vs. 14 seconds) while achieving the same throughput (1.2 rps), as shown in Figure 2. These trends hold when evaluating Llama-2 13B as shown in Figure 3.

\subsection{Effective Throughput Analysis}

Under the effective throughput analysis that considers both first token latency and the rate at which generation occurs, DeepSpeed-FastGen provides up to 2.3x higher throughput than vLLM. Figure 4 presents a comparative analysis of the effective throughputs of DeepSpeed-FastGen and vLLM. Each plotted point denotes the effective throughput derived from a specific number of clients. As we scaled the number of clients, we initially observed an increase in effective throughput. However, the latency also significantly increases as the number of clients approaches the system's capacity, causing many requests to fail in meeting the SLA. Consequently, the effective throughput will either saturate or decrease at some point. From a usability perspective, it's not particularly relevant how many clients are required to achieve the max effective throughput; the maximum point of the line is the optimal serving point.

\begin{figure}
\centering
\includegraphics[width=\textwidth]{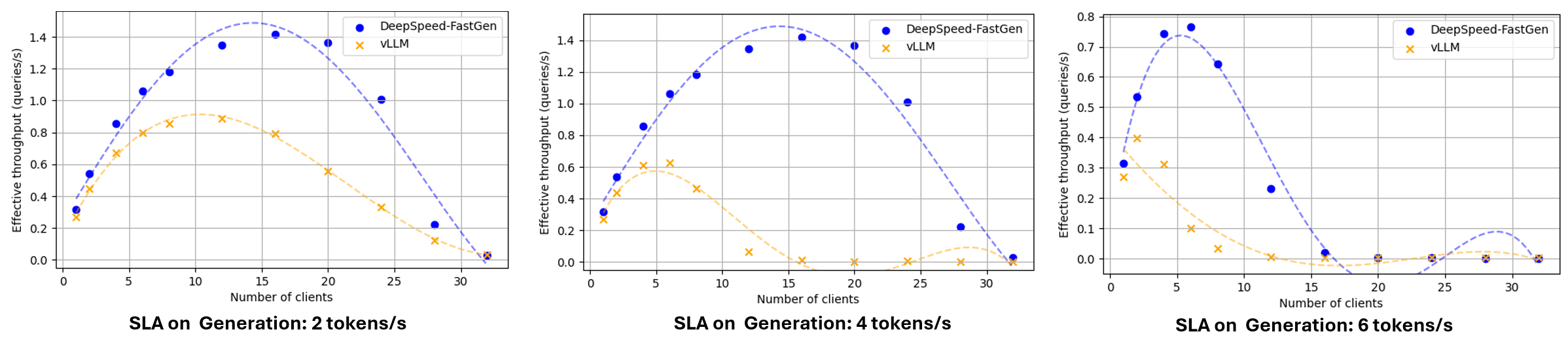}
\caption{Effective throughput of DeepSpeed-FastGen and vLLM (Llama 2 70B/A100-80GB using tensor parallelism across 4 A100-80GB GPUs. A normal distribution was applied to prompt and generation lengths with averages of 2600 and 60, respectively, and a 30\% variance).}
\label{fig:effective_throughput}
\end{figure}

\subsection{Significant Tail Latency Reduction for Token Generation}

Figure 5 displays the P50, P90, and P95 latencies of the generation processes. Both vLLM and DeepSpeed-FastGen exhibit similar P50 latencies, but vLLM demonstrates significantly higher latencies for P90 and P95. Regarding the P95 latencies, DeepSpeed-FastGen achieved a reduction of 3.7 times.

This discrepancy is due to a noticeable spike in vLLM's generation latency when it preempts the ongoing generation to process new prompts.
In contrast, DeepSpeed-FastGen typically processes the prompt and generation for previous requests concurrently, leading to much more consistent generation latency.

\subsection{Scalability using Load Balancing}

DeepSpeed-FastGen offers replica-level load balancing that evenly distributes requests across multiple servers, allowing you to effortlessly scale up your application. Figure 6 illustrates the scalability of DeepSpeed-FastGen when employing the load balancer and up to 16 replicas. Note that we utilized 4 A100 GPUs to compute the Llama 2 70B model. In total, we employed 8 nodes to run the 16 replicas. The results demonstrate nearly perfect scalability with DeepSpeed-FastGen.
Given that the throughput of a single replica is 1.46 queries/sec, the throughput with 16 replicas reaches 23.7 queries/sec, marking a linear 16x increase compared to a single replica.

\subsection{Other Hardware Platforms}

In addition to the deep analysis on A100, we provide additional benchmarking results for H100 and A6000. The same performance trends were observed on both A6000 and H100 as A100.

\begin{figure}[htbp!]
\centering
\includegraphics[width=0.5\textwidth]{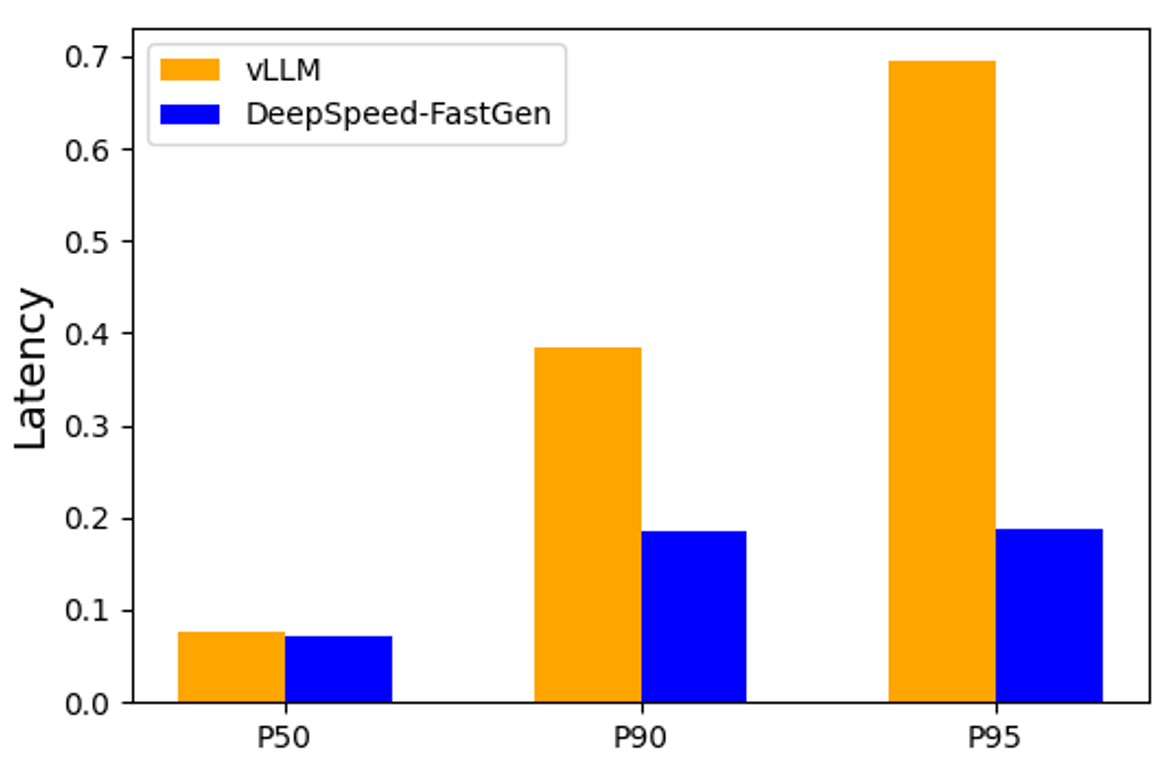}
\caption{Per-Token generation Latency of Llama 2 70B/A100-80GB using tensor parallelism across 4 A100-80GB GPUs, 16 clients. A normal distribution was applied to prompt and generation lengths with averages of 2600 and 128, respectively, and a 30\% variance.}
\label{fig:token_latency}
\end{figure}

\begin{figure}[htbp!]
\centering
\includegraphics[width=0.5\textwidth]{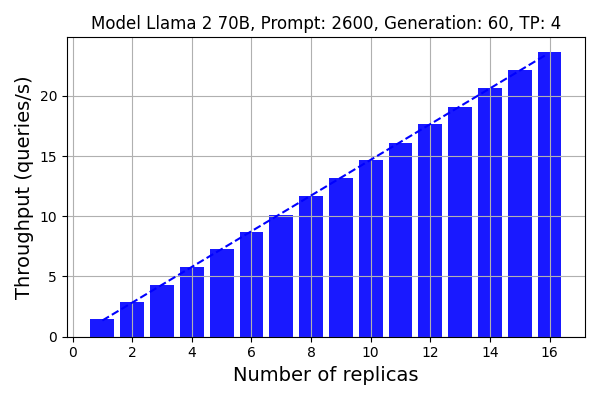}
\caption{Scalability using the load balancing feature. A normal distribution was applied to prompt and generation lengths with averages of 2600 and 60, respectively, and a 30\% variance.}
\label{fig:repl_scale_llama70b_tp4_p2600g60}
\end{figure}

\begin{figure}[htbp!]
\centering
\includegraphics[width=0.8\textwidth]{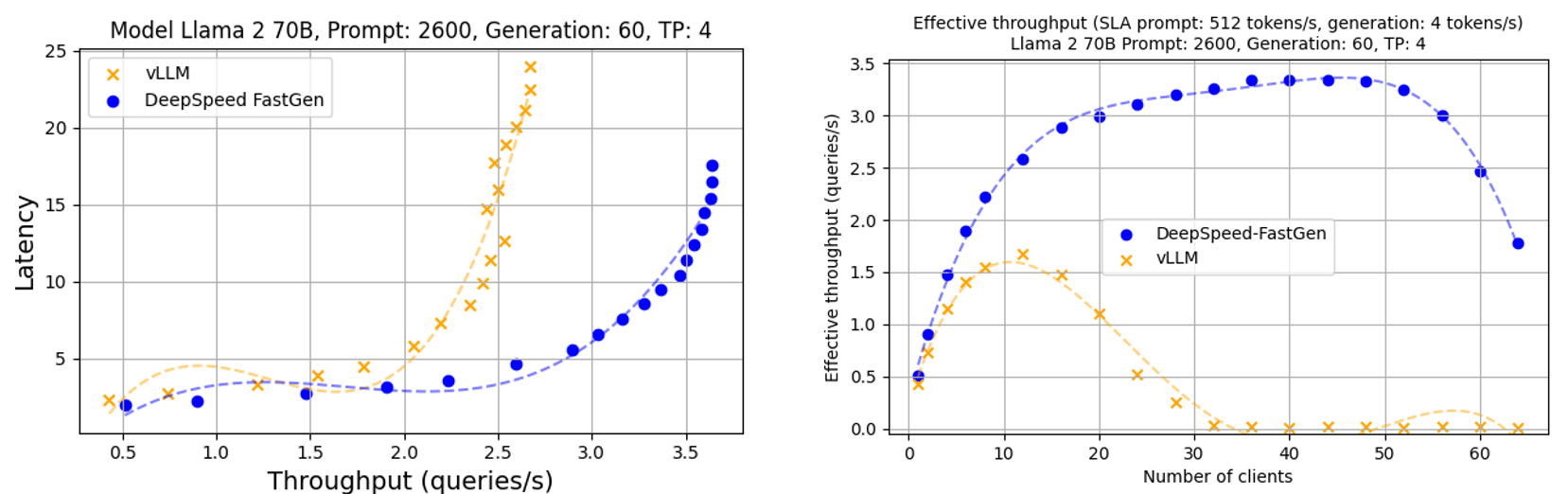}
\caption{Throughput-latency curve and effective throughput of Llama 2 70b using 8 H100 GPUs. A normal distribution was applied to prompt and generation lengths with averages of 2600 and 60, respectively, and a 30\% variance.}
\label{fig:H100_benchmark}
\end{figure}

\begin{figure}[htbp!]
\centering
\includegraphics[width=0.7\textwidth]{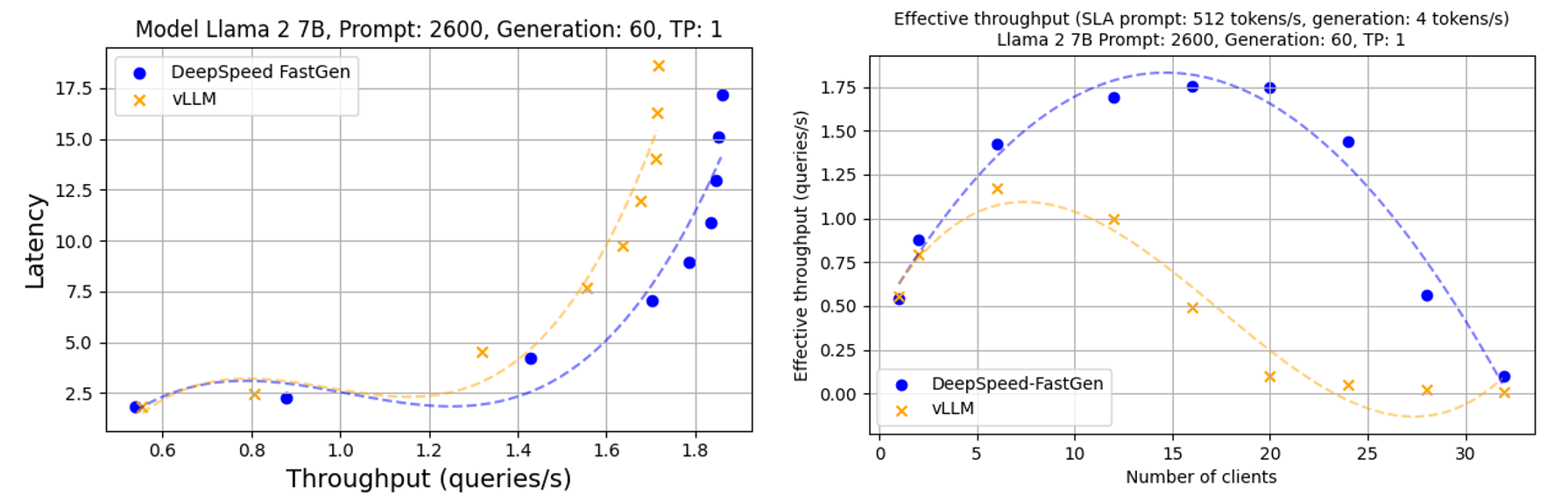}
\caption{Throughput-latency curve and effective throughput of Llama 2 7b using A6000. A normal distribution was applied to prompt and generation lengths with averages of 2600 and 60, respectively, and a 30\% variance.}
\label{fig:A6000_benchmark}
\end{figure}

\section{DeepSpeed-FastGen: Implementation and Usage 
}
\label{sec:implementation}

DeepSpeed-FastGen is the synergistic composition of
\href{https://github.com/microsoft/DeepSpeed-MII}{DeepSpeed-MII}
and
\href{https://github.com/microsoft/DeepSpeed}{DeepSpeed-Inference}
as illustrated in the figure below. Together, both of these software packages provide various components of the system including the frontend APIs, the host and device infrastructure to schedule batches using Dynamic SplitFuse, optimized kernel implementations, and the tools to construct new model implementations.

\begin{figure}[htbp!]
\centering
\includegraphics[width=0.9\textwidth]{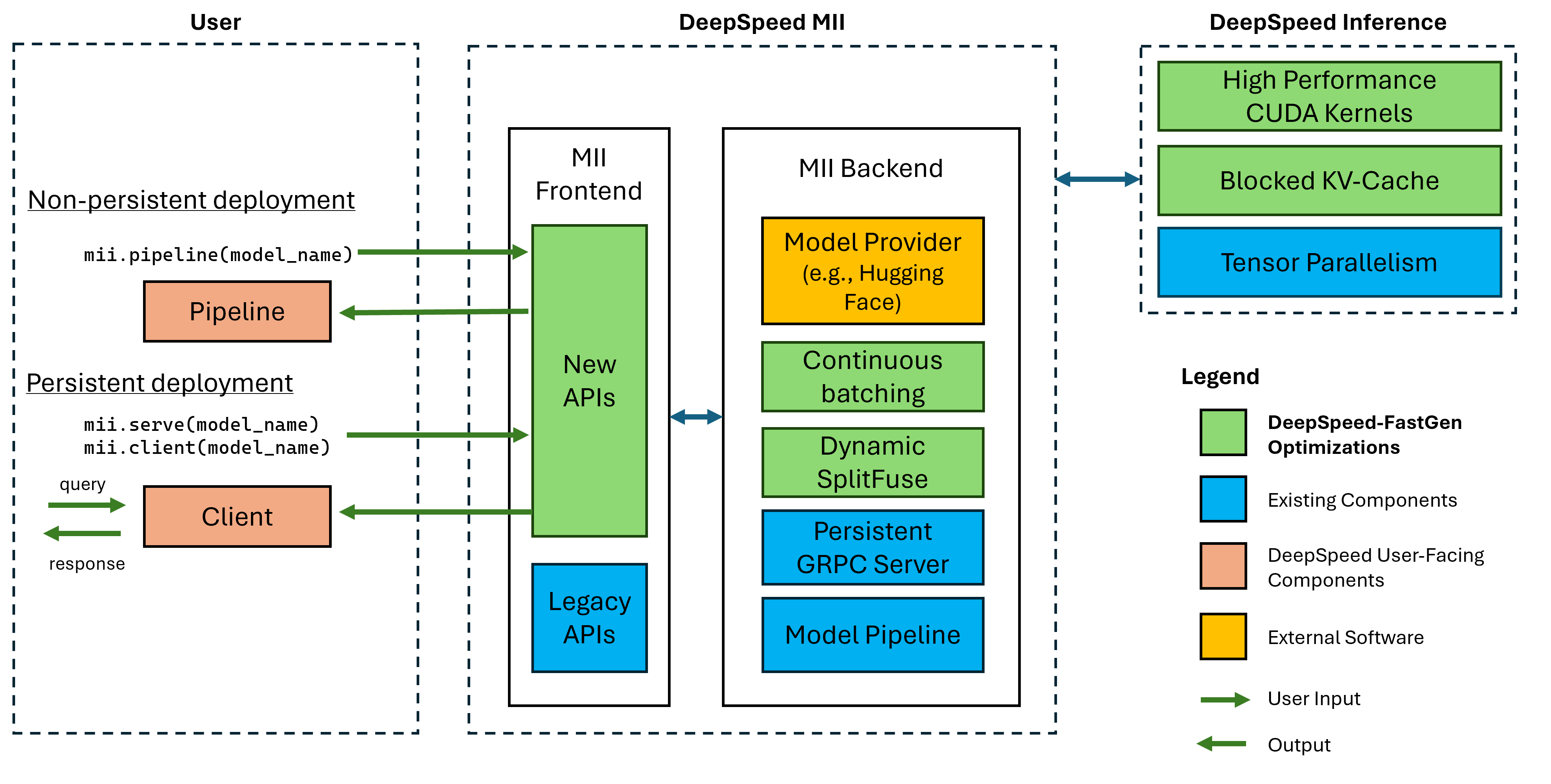}
\caption{Architecture of DeepSpeed-FastGen}
\label{fig:fastgen-arch-light}
\end{figure}

The fastest way to get started with our alpha release of DeepSpeed-FastGen is by running the following command:

\begin{lstlisting}[language=Python]
pip install deepspeed-mii
\end{lstlisting}


\clearpage
\subsection{Supported Models}

We currently support the following HuggingFace model families\footnote{\url{https://huggingface.co/models}} in this alpha release of DeepSpeed-FastGen:

\begin{itemize}
\item LLaMA and LLaMA-2
\item Mistral
\item Facebook OPT
\end{itemize}

All current models leverage HuggingFace APIs in our backend to provide both the model weights and the model's corresponding tokenizer. We plan to add additional models in the coming weeks and months after the initial release. If there are specific model architectures you would like supported, please
\href{https://github.com/microsoft/DeepSpeed-MII/issues}{file an issue}
and let us know.

\subsection{Deployment Options}

All of the examples below are runnable in 
\href{https://github.com/microsoft/DeepSpeedExamples/tree/master/inference/mii}{DeepSpeedExamples}. Once installed you have two options for deployment: an interactive non-persistent pipeline or a persistent serving deployment:

\subsubsection{Non-persistent Pipeline}

The non-persistent pipeline deployment is a great and fast way to get started and can be done with only a few lines of code. Non-persistent models are only around for the duration of the python script you are running but are useful for temporary interactive sessions.

\begin{lstlisting}[language=Python]
from mii import pipeline
pipe = pipeline("mistralai/Mistral-7B-v0.1")
output = pipe(["Hello, my name is", "DeepSpeed is"], max_new_tokens=128)
print(output)
\end{lstlisting}

\subsubsection{Persistent Deployment}

A persistent deployment is ideal for use with long-running and production applications. The persistent deployment uses a lightweight GRPC server that can be created using the following 2 lines:

\begin{lstlisting}[language=Python]
import mii
mii.serve("mistralai/Mistral-7B-v0.1")
\end{lstlisting}

The above server can be queried by multiple clients at once thanks to the built-in load balancer from DeepSpeed-MII. Creating a client also just takes 2 lines of code:

\begin{lstlisting}[language=Python]
client = mii.client("mistralai/Mistral-7B-v0.1")
output = client.generate("Deepspeed is", max_new_tokens=128)
print(output)
\end{lstlisting}

A persistent deployment can be terminated when it is no longer needed:

\begin{lstlisting}[language=Python]
client.terminate_server()
\end{lstlisting}

\subsection{Advanced Installation Information}

For ease of use and a significant reduction in lengthy compile times that many projects require in this space, we distribute a pre-compiled Python wheel covering the majority of our custom kernels through a new library called \href{https://github.com/microsoft/DeepSpeed-Kernels}{DeepSpeed-Kernels}. We have found this library to be very portable across environments with NVIDIA GPUs with compute capabilities 8.0+ (Ampere+), CUDA 11.6+, and Ubuntu 20+. In most cases, you shouldn't even need to know this library exists as it is a dependency of DeepSpeed-MII and will be installed with it. However, if for whatever reason you need to compile our kernels manually please see our \href{https://github.com/microsoft/DeepSpeed-Kernels\#source}{advanced installation docs}.

\section{Release: Try Out DeepSpeed-FastGen}

We are very excited to share this DeepSpeed-FastGen alpha release. To get started, please visit our \href{https://github.com/microsoft/DeepSpeed-MII}{GitHub Landing Page} page for DeepSpeed-MII.

DeepSpeed-FastGen is part of the bigger DeepSpeed ecosystem comprising a multitude of Deep Learning systems and modeling technologies. To learn more,

\begin{itemize}
    \item Please visit our \href{https://www.deepspeed.ai/}{website} for detailed blog posts, tutorials, and helpful documentation.
    \item You can also follow us on our \href{https://twitter.com/MSFTDeepSpeed}{English Twitter}, \href{https://twitter.com/MSFTDeepSpeedJP}{Japanese Twitter}, and \href{https://www.zhihu.com/people/deepspeed}{Chinese Zhihu} for the latest news on DeepSpeed.
\end{itemize}

DeepSpeed welcomes your contributions! We encourage you to report issues, contribute PRs, and join discussions on the \href{https://github.com/microsoft/DeepSpeed/}{DeepSpeed GitHub} page. Please see our \href{https://github.com/microsoft/DeepSpeed/blob/master/CONTRIBUTING.md}{contributing guide} for more details. We are open to collaborations with universities, research labs, and companies, such as those working together on deep learning research, applying DeepSpeed to empower real-world AI models and applications, and so on. For such requests (and other requests unsuitable for GitHub), please directly email to \href{mailto:deepspeed-info@microsoft.com}{deepspeed-info@microsoft.com}.

\subsection{Roadmap}
The following items are on our roadmap and we plan to engage with our community on these through our GitHub issues and PRs:

\begin{itemize}
    \item Performance improvements
    \item Broader model support
    \item New hardware backends through collaboration with partners
    \item Release performance benchmarks (used to generate plots in this blog)
\end{itemize}

\textbf{``Star'' our \href{https://github.com/microsoft/DeepSpeed/}{DeepSpeed GitHub} and \href{https://github.com/microsoft/DeepSpeed-MII/}{DeepSpeedMII GitHub} repositories if you like our work!}

\section*{Acknowledgment}
We thank the entire DeepSpeed team for their contributions on developing, debugging, testing, and releasing the DeepSpeed-FastGen software. We would like to thank various open-source community projects including HuggingFace, vLLM, and HuggingFace TGI. We have leveraged HF APIs to support models and tokenizers in our alpha release and will continue to add more models. We especially acknowledge and thank the developers of \href{https://github.com/Dao-AILab/flash-attention}{Flash Attention}\cite{dao2022flashattention,dao2023flashattention2} for their great work. We have extensively leveraged FlashAttention kernels in our system with modifications that have been acknowledged in our code repositories at appropriate file headers. Finally, we want to thank the developers of \href{https://github.com/NVIDIA/FasterTransformer}{FasterTransformer}~\cite{fastertransformer} kernels that we have used in our MoE kernels (released as part of DeepSpeed-Kernels repository).

\bibliographystyle{unsrt}
\bibliography{references}

\newpage
\appendix

\end{document}